# DIGITAL IMAGE CORRELATION: FROM DISPLACEMENT MEASUREMENT TO IDENTIFICATION OF ELASTIC PROPERTIES

François H<small>ILD</small>[1,*] and Stéphane R<small>OUX</small>[2]

[1]*LMT-Cachan, ENS-Cachan / CNRS UMR 8535 / Université Paris 6*

*61 avenue du Président Wilson, F-94235 Cachan Cedex, France*

[2]*Laboratoire Surface du Verre et Interfaces, UMR 125 CNRS / Saint Gobain*

*39 quai Lucien Lefranc, F-93303 Aubervilliers Cedex, France*

**Running head title**: DIC: from measurement to identification

[*]Corresponding author: Fax: +33 1 47 40 22 40, Email: hild@lmt.ens-cachan.fr

# DIGITAL IMAGE CORRELATION:

# FROM DISPLACEMENT MEASUREMENT

# TO IDENTIFICATION OF ELASTIC PROPERTIES

François HILD and Stéphane ROUX

**Abstract:**

The current development of digital image correlation, whose displacement uncertainty is well below the pixel value, enables one to better characterise the behaviour of materials and the response of structures to external loads. A general presentation of the extraction of displacement fields from the knowledge of pictures taken at different instants of an experiment is given. Different strategies can be followed to achieve a sub-pixel uncertainty. From these measurements, new identification procedures are devised making use of full-field measures. A priori or a posteriori routes can be followed. They are illustrated on the analysis of a Brazilian test.

**Keywords:**

Brazilian test, correlation algorithm, integrated approach, kinematic measurements, Photomechanics.

# 1. Introduction

With the development of complex thermomechanical experiments, *full-field* displacement and strain measurements are often needed. In recent years, the development of robust, fast and affordable tools for full-field kinematic measurements prompts more and more experimentalists to resort to such techniques. The latter can be adapted to evaluate 2D or 3D displacement fields of the surface of solids [1] and more recently even in the bulk of a solid [2]. More generally, full-field measurements constitute an opportunity to bridge the gap between experiments and simulations allowing for direct displacement and strain comparisons.

Full-field kinematic measures can be used in a variety of ways, namely:

- to check boundary conditions before performing the mechanical test itself [3]. In that case, it allows the experimentalist to control whether the boundary conditions correspond to the desired ones;

- to control an experiment by using optical means as opposed to gauges or extensometers [4,5]. In this area, one can envision using hybrid techniques combining FE simulations as input "signals" to be compared with actual full-field measurements as is now performed for so-called pseudo-dynamic experiments [6,7,8,9];

- to perform heterogeneous tests on materials [10] or structures [11] for which single measurements (*e.g.,* by strain gauges, extensometers, clip gauges) are not sufficient to fully monitor an experiment, and particularly when the spatial heterogeneity is not known *a priori* (*e.g.,* strain localisation [12,13], damage localisation [14] or crack initiation and propagation [15,16]);

- to study an experiment by using contactless techniques. This provides useful solutions to aggressive, hot, corrosive environments, or very soft solids for which gauges are not adapted (*e.g.,* polymers [17], wood and paper [18], mineral wool [19]);

- to identify material properties or validate constitutive models [20,21,22,23].

On the one hand, the experiments can be performed with a controlled illumination device (i.e., laser) for which holographic or shearing interferometry, ESPI and laser speckle photography are some possibilities [24,1]. These techniques often require careful preparations and are not always well suited for experiments in a conventional testing lab. On the other hand, white light can also be used in fringe projection methods as well as moiré or speckle photography. For the latter, by acquiring pictures for different load levels, it is possible to determine the in plane or 3D displacement field by matching different zones of two pictures.

Section 2 is devoted to the presentation of the general principles of a correlation technique as a measurement means of kinematic data. The basic tools of the method are introduced and the displacement uncertainty is assessed. Section 3 discusses different *a posteriori* strategies to identify material parameters. An application of the digital image correlation technique is discussed to determine the elastic properties of polycarbonate by analysing a Brazilian test. In Section 4, an *integrated* approach is presented. It consists in a *coupled* and *a priori* way of dealing with measurement *and* identification. It is also applied to the analysis of the same Brazilian test.

## 2. Kinematic measurements by using DIC

Digital image correlation was proposed at the beginning of the 1980's [25,26] when applied in solid mechanics. Many of the procedures available nowadays are based on Correlation Image Velocimetry (CIV) methods [27] directly inspired from earlier developments in fluid mechanics [28,29].

In the present section, the displacements are expressed in terms of pixels, the only quantity at hand when pictures are analysed. The physical scale can vary from micrometers [16] to meters [11] depending upon the magnification means used between the

CCD sensor and the observed surface. It is worth noting that the physical scale has a negligible impact on the correlation results, provided a random texture is obtained whose typical scale (i.e., correlation radius) is of the order of a few pixels.

### 2.1. Correlation principle

Let us consider a reference image, defined as $f$, e.g., a gray level distribution. An in-plane displacement field $\mathbf{u}(\mathbf{x})$ is defined. The passive advection of the texture $f$ by the displacement field defines a "deformed image," $g$, such that

$$g(\mathbf{x}+\mathbf{u}) = f(\mathbf{x}) + b(\mathbf{x}) \qquad (1)$$

where $b(.)$ is noise induced by image acquisition. In the following, it is assumed that the noise level, $b$, can be neglected either because of its low amplitude with respect to those of $f$ and $g$, or because of its scale separation with significant components of the displacement field. Equation (1) is the integral form of the "optical flow equation". The problem to address is the determination of the displacement field $\mathbf{u}$ from the exclusive knowledge of $f$ and $g$. As such, the problem is ill-posed, unless additional assumptions are made on the regularity of the displacement field so that the information is sufficient to determine $\mathbf{u}$ with a reasonable accuracy [30]. Let us introduce the following objective functional $\Phi$ operating on displacement fields $\mathbf{v}$

$$\Phi(\mathbf{v}) = \iint [g(\mathbf{x}+\mathbf{v}) - f(\mathbf{x})]^2 d\mathbf{x} \qquad (2)$$

In the absence of noise, this functional reaches its minimum value, 0, for $\mathbf{v} = \mathbf{u}$ [see equation (1)]. The trial displacement $\mathbf{v}$ can be any choice as will be discussed below. To fulfil a smoothness assumption on $\mathbf{u}$, $\mathbf{v}$ can be low-pass filtered or chosen in a subspace of suited functions. Let us first assume that $f$ and $g$ are sufficiently smooth on small scales, and the displacement small enough in amplitude so that a Taylor expansion of $g$ up to the first order can be introduced in equation (2)

$$\Phi(\mathbf{v}) = \iint [g(\mathbf{x}) - f(\mathbf{x}) + \mathbf{v}(\mathbf{x}).\nabla g(\mathbf{x})]^2 d\mathbf{x} \qquad (3)$$

Equation (3) corresponds to the objective functional associated to the optical flow equation. The measurement of a displacement field is therefore an ill-posed problem. The displacement can only be measured along the direction of the intensity gradient. Consequently, additional hypotheses have to be proposed to solve the problem. A spatial regularisation was introduced by Horn and Schunck [30] and consists in a looking for *smooth* displacement solutions. However, this method is not appropriate for problems dealing with discontinuities in the apparent displacement [31]. In the latter case, the quadratic penalisation is replaced by "smoother" penalisations based on robust statistics [32,33,34]. Another strategy consists in decomposing the displacement field as a linear combination of basis functions $\eta_i(\mathbf{x})$

$$\mathbf{v} = \sum_i \mathbf{v}_i \eta_i(\mathbf{x}) \qquad (4)$$

so that $\Phi$ becomes a quadratic form in the (vector) amplitudes $\mathbf{v}_i$. The extremality condition thus dictates, for all $j$,

$$\left[\iint (\nabla g \otimes \nabla g)(\mathbf{x}) : (\eta_j \otimes \eta_k)(\mathbf{x}) d\mathbf{x}\right] \mathbf{v}_k = \iint [f(\mathbf{x}) - g(\mathbf{x})] \nabla g(\mathbf{x}).\eta_j(\mathbf{x}) d\mathbf{x} \qquad (5)$$

where $\otimes$ denotes the dyadic product and : the contraction with respect to two indices. This system can be written in a matrix form as

$$[\mathbf{M}]\{\mathbf{w}\} = \{\mathbf{m}\} \qquad (6)$$

where $\{\mathbf{w}\}$ is a vector containing all the unknown components $\mathbf{v}_i$, $[\mathbf{M}]$ and $\{\mathbf{m}\}$ are *known* quantities dependent upon $f$, $g$, and $\eta$. One can note that the dyadic product $\nabla g \otimes \nabla g$ by itself cannot be inverted (it has always a zero eigenvalue in the direction normal to the gradient of $g$), and hence this formula cannot be used to determine $\mathbf{v}$, if $\eta_i$ tends to a Dirac distribution, as anticipated from the remark on the necessary regularity of $\mathbf{v}$. However, if the functions $\eta_i$ are chosen to be restricted to wavelengths much greater than the correlation length of the texture, the left-hand side operator becomes a *genuine* definite positive operator, and the product with $\eta_j \otimes \eta_k$ is nothing but a filtering of the operator $\nabla g \otimes \nabla g$. In Section 4, an example will be discussed in which the basis functions are associated to the exact displacement solution of a Brazilian test.

As a particular case, one can note that if **v** is simply a rigid body translation, then the problem can be addressed by using the standard cross-correlation technique [25,26] to maximise $\Phi$ in equation (2)

$$(f \bullet g)(\mathbf{v}) = \iint f(\mathbf{x}) g(\mathbf{x} + \mathbf{v}) d\mathbf{x} \qquad (7)$$

where $\bullet$ denotes the cross-correlation product. The computation of a cross-correlation can be performed either in the original space [26,35] or in the Fourier space [36,37,38] by using fast Fourier transforms (FFT)

$$f \bullet g = l \, \mathrm{FFT}^{-1}\left[\overline{\mathrm{FFT}(f)}\, \mathrm{FFT}(g)\right] \qquad (8)$$

where the complex conjugate is overlined, and the correlation domain has $l \times l$ samples. Another useful result associated to Fourier transforms is the shift / modulation property. Let us consider the shift operator $T_\mathbf{d}$ defined by

$$T_\mathbf{d} f(\mathbf{x}) = f(\mathbf{x} + \mathbf{d}) \qquad (9)$$

where **d** is the shift parameter. The FFT of $T_\mathbf{d}$ becomes

$$\mathrm{FFT}[E_\mathbf{d} f] = E_\mathbf{d}\, \mathrm{FFT}[f] \qquad (10)$$

where the modulation operator $E_\mathbf{d}$ is defined by

$$E_\mathbf{d} \hat{f}(\mathbf{k}) = \exp(-2i\pi \mathbf{k}.\mathbf{d})\, \hat{f}(\mathbf{k}) \qquad (11)$$

In practice two images are considered. The first one, referred to as the "reference image" and the second one, called the "deformed image." The user usually chooses the size of the zones of interest (ZOI) by setting the size $l$ so that $l \times l$ pixels are considered. To map the whole image, the second parameter to choose is the separation $\delta$ between two consecutive

ZOIs. The latter defines the mesh formed by the centres of each ZOI used to analyse the displacement field (figure 1). A first step usually consists in finding displacements as pixel integers by seeking the maximum of the correlation function in the neighbourhood, whose size can be larger than the ZOI size, of the initial ZOI location. Conversely, coarse-graining techniques can be used to get this first information [19]. Let us stress the importance of this iterative multiscale procedure. In particular, following the route of equation (5), the adequacy of the gradient operator is suited for small displacement amplitudes, small is understood here in comparison with either the pixel size of the shortest wavelength present in an image after low-pass filtering. In this multiscale procedure, the philosophy is to first determine the largest displacements by using a coarse or filtered description of the two images, and progressively restoring details on images where the previous displacement field has been corrected for. This procedure guarantees both the robustness and accuracy of the results. The next step is to evaluate the sub-pixel correction. This is needed in many applications when small strains are suspected to occur.

### *2.2. Sub-pixel algorithms*

*2.2.1. Locally continuous displacements*

To determine a sub-pixel displacement, interpolations are needed to estimate $g(\mathbf{x}+\mathbf{v})$ in equation (7) at non-integer values. Different choices can be made to interpolate the texture, namely, by using bilinear, bicubic, or spline functions. It has a direct impact on the performance of the correlation algorithm [39]. The next options are related to the optimisation algorithm (e.g., first gradient descent, Newton-Raphson, Levenberg-Marquard) and the sought displacements $\mathbf{v}(\mathbf{x})$ for the considered ZOI (e.g., translation, translation + rotation, translation + rotation + uniform strain) [40].

Another way of obtaining a sub-pixel displacement correction $\boldsymbol{\delta v}$ is given by the determination of the maximum of a parabolic interpolation of the correlation function. The interpolation is performed by considering the maximum pixel and its eight nearest neighbours. Therefore, one obtains a sub-pixel value. By using the shift / modulation property [equation (10)], one can move the windowed ZOI in the deformed image by an amount $\delta\mathbf{v}$. Since an interpolation is used, one may induce some errors requiring to re-iterate by considering the new ZOI until a convergence criterion is reached. The criterion checks whether the maximum of the interpolated correlation function increases as the number of iteration increases. Otherwise, the iteration scheme is stopped [10].

*2.2.2. Globally continuous displacements: Q1 formulation*

The general method presented in equations (4-6) can be applied to a large variety of test functions. For illustration purposes, Q1-elements defined on a square grid (bi-linear functions of *x* and *y*) are chosen here. In contrast to the previous approach, the displacement field is now *globally* continuous as in FE computations. Consequently, there is only one correlation parameter defined by the element size *l*. As previously discussed, the very first step is a determination of the global translation performed on coarse images, so as to capture the large amplitudes of the displacement field, and progressively finer details are restored. This linear system naturally leads to subpixel displacements, which can be corrected for by using the shift / modulation properties of FFTs.

**2.3. Displacement uncertainty and resolution**

There are different ways of evaluating a correlation algorithm. The first one consists in creating artificial pictures and applying known displacements to different "markers" of the picture. Since all the conditions are known, it is possible to evaluate the error in terms of average, standard deviation. However, these artificial textures are not always representative of real situations. The second one consists in recording pictures prior to an experiment and applying rigid body motions. The displacement levels are not known. Yet the standard displacement and strain uncertainties can be estimated and should be as low as possible. The latter are usually lower bounds to the actual values. The third route consists in recording a picture of the considered experiment and applying artificially known displacements. It allows one to estimate the performances associated to actual pictures considered in experiments. This last case is now considered. In the following, only displacements are going to be considered. The uncertainty of the latter will be estimated when the correlation parameters are modified. To have *independent* estimates of the displacements in the following analyses, the separation $\delta$ is equal to the ZOI size $l$.

A constant displacement varying between 0 and 1 pixel, with an increment of 0.1 pixel is applied artificially by using the shift / modulation property of Fourier transforms. In the present study, the reference picture of figure 1 is considered. Figure 2a shows the standard displacement uncertainty as a function of the applied displacement for different ZOI sizes $l$ of the reference picture shown in figure 1a. For a given ZOI size, the maximum uncertainty is reached when the prescribed displacement is equal to 0.5 pixel. In this case, the information between each ZOI in the reference and deformed pictures is the most biased. The corresponding displacement uncertainty $\sigma_u$, defined as the mean of the standard displacement uncertainties, is plotted as a function of the ZOI size (figure 2b). A power law with an exponent $\alpha$ of the order 0.59 is obtained for the displacement uncertainty $\sigma_u$

$$\sigma_u = \frac{A^{\alpha+1}}{l^\alpha} \tag{12}$$

with $A = 0.48$ pixel, thereby indicating that the displacement uncertainty and the spatial resolution are the result of a compromise.

The same procedure can be applied to the determination of Q1-displacement fields. One notes a certain similarity of shape in the standard displacement uncertainty, with a pronounced peak for a displacement equal to 0.5 pixel, and a lower uncertainty for large element sizes. However, it is also worth noting that in contrast with classical DIC, the uncertainty is not a periodic function with a period of one pixel. For large displacements, and without any specific care for handling "large" displacements (such as the multiresolution scheme presented above or a first global correction), the 1$^{st}$ order Taylor expansion fails to capture the proper correction. Apart from this first difference, the Q1-displacement interpolation tends to give more accurate results than the direct DIC technique, especially when the ZOI sizes are large. The mean displacement uncertainty follows the power law (12), with an exponent $\alpha = 1.40$ and a prefactor $A = 0.88$ pixel by using the full dynamic range of the pictures (i.e., 12 bits). The requirement of global displacement continuity appears to be the key in reducing the displacement uncertainty as well as a good dynamic range for large ZOI sizes when compared to FFT-DIC results. Let us note that the poor texture (in terms of variations compared with the dynamic range of the CCD camera as can be seen in figure 1), making it a difficult test case, leads to an overall performance not as good as that observed in other situations [16,13].

## 2.4. Analysis of a Brazilian test

A so-called Brazilian test [41,42] is analysed with a circular disk 155 mm in diameter, and 10 mm in thickness made of polycarbonate and submitted to diametrical compression (figure 3). The disk is sprayed with black and white paint. As discussed above, the gray level distribution is considered as very difficult since the white texture is not very important (figure 1). The previous uncertainty analysis was performed with the reference picture of the present experiment. The physical size of one pixel is 98 μm.

Figure 4 shows the displacement field obtained with a Q1-procedure and a size $l = 16$ pixels. It can be noted that the displacement variations remain small and most of the local fluctuations can be attributed to the uncertainty associated to the chosen correlation parameter. For the vertical displacements, the measured range is less than 5 pixels (i.e., less than 0.5 mm). Conversely, for horizontal displacements, the range is of the order of 2.5 pixels

(i.e., less than 0.25 mm) with no significant increase in terms of noise associated to the displacement uncertainty (of the order of 0.02 pixel or less than 2 µm). Since the Q1-formulation is based upon equation (3), the gray level residuals defined by $\sqrt{\Phi(\mathbf{v})}$ can be computed for each pixel. Figure 5 shows the map of gray level residuals. An RMS value of 18 gray levels is obtained to be compared with the dynamic range of the pictures equal to 12 bits (i.e., 4096 gray levels). Apart from few uncorrelated high residuals, their fluctuations are very small.

Most of the above discussion is based on the displacement field itself, i.e., the measured quantity. In practice, one is often more interested in strains rather than displacements *per se*. However, being a derivative, the strain is much more sensitive to the high frequency noise, which naturally goes together with any determination. Figure 6 shows the normal strain field in the direction perpendicular to the compression axis, which controls the diametrical fracture typically observed in this test [41,42]. It is observed that this map is dominated by the measurement uncertainty. This is explained by the fact that the strain uncertainty $\sigma_\varepsilon$ scales as $\sqrt{2}\sigma_u/l$ [13] so that a first order estimate is given by $\sigma_\varepsilon \approx 1.8 \times 10^{-3}$ when using the results of figure 2b with $l = 16$ pixels. Consequently, the strain uncertainty is greater than the mean value (here $1.5 \times 10^{-3}$).

## 3. Identification techniques using full-field measurement techniques

Full-field measurement techniques are more and more used to analyse experiments. Current developments aim at devising identification strategies making use of large amount of measurement data. The interested reader may refer to a recent review on the subject in elasticity [43].

A first approach is based on minimising, in the same spirit as in Section 2 for the measurement stage, displacements either determined from closed-form solutions (e.g.,

tension, bending, cracked samples, Flamant problem [44], Brazilian test) or from FE simulation [45]

$$\Theta(\alpha) = \sum \|\mathbf{v}_{Id}(\alpha) - \mathbf{v}_{me}\|^2 \qquad (13)$$

where $\alpha$ is a set of unknown parameters. The norm $\|.\|$ can be the classical norm 2 [46] or based upon a constitutive equation error [47,48]. The latter was used in a variety of situations [49,50,51,23]. It is worth noting that FE updating usually requires numerous simulations and the load levels can also be added as an input [52]. Conversely, when closed-form solutions can be used and when the displacements are linearly related to the unknown parameters, a linear system is obtained [44].

Other identification routes can also be followed. The so-called virtual field method has been used to identify homogeneous properties of composites [20,53,54] (i.e., in anisotropic elasticity). Another procedure is based upon the reciprocity gap [55] that can also be used to determine the local elastic field or detect cracks in elastic media [56]. An alternative method solely based on displacement field data can also be used [22,57] to identify elastic property fields and damage fields through the elasticity/damage state coupling.

In the present case, the simple approach given by equation (13) is followed and applied to determine the elastic parameters of the polycarbonate used in the Brazilian test described in Section 2.4. This test case is considered as very difficult since the displacement uncertainties are quite high. To analyse the Brazilian test, a closed-form solution for the displacements can be derived from the Kolossov-Muskhelishvili potentials $\varphi$ and $\psi$ [58]

$$2\mu\mathbf{v} = \kappa\varphi(\mathbf{z}) - \mathbf{z}\overline{\varphi'(\mathbf{z})} - \overline{\psi(\mathbf{z})} \qquad (14)$$

where $\mu$ is the Lamé's shear modulus, $\kappa$ a dimensionless elastic coefficient related to the Poisson's ratio $\nu$ according to $(3 - \nu)/(1 + \nu)$ for plane stress conditions, $\mathbf{z} = x + iy$, $\mathbf{v} = v_x + iv_y$,

$$\varphi(\mathbf{z}) = \frac{FR}{2\pi}\left[\frac{\mathbf{z}}{R} - \log\left(\frac{R+\mathbf{z}}{R-\mathbf{z}}\right)\right] , \quad \psi(\mathbf{z}) = \frac{FR}{2\pi}\left[\frac{R}{R-\mathbf{z}} - \frac{R}{R+\mathbf{z}} + \log\left(\frac{R+\mathbf{z}}{R-\mathbf{z}}\right)\right] \quad (15)$$

and $R$ is the disk radius. Equation (10) can be rewritten as

$$\Theta(\mu,\kappa) = \sum_n |A_n\mu + B_n\kappa + C_n|^2 \quad (16)$$

with

$$A_n = 2\mathbf{v}_{mes}(\mathbf{z}_n), \quad B_n = -\varphi(\mathbf{z}_n), \quad C_n = \mathbf{z}_n\overline{\varphi'(\mathbf{z}_n)} + \overline{\psi(\mathbf{z}_n)} \quad (17)$$

where $\mathbf{z}_n$ denotes the location of the measurement points. To minimise $\Theta$ in Equation (16), one has to solve a $2 \times 2$ linear system in the unknowns $\mu$ and $\kappa$ [44]

$$\sum_n \begin{bmatrix} |A_n|^2 & \Re(A_n \overline{B}_n) \\ \Re(A_n \overline{B}_n) & |B_n|^2 \end{bmatrix} \begin{Bmatrix} \mu \\ \kappa \end{Bmatrix} = \sum_n \begin{Bmatrix} \Re(A_n \overline{C}_n) \\ \Re(B_n \overline{C}_n) \end{Bmatrix} \quad (18)$$

When using data obtained by classical DIC, the best match for the displacements is given when $\mu = 859$ MPa and $\kappa = 1.70$ (i.e., $E = 2.54$ GPa and $\nu = 0.48$) for a ZOI size $l = 64$ pixels (i.e., the displacement uncertainty is less than 0.03 pixel or less than 3 µm). The same analysis is performed with the displacements measured with the Q1-DIC algorithm (figure 4). For an element size of 16 pixels, $\mu = 846$ MPa and $\kappa = 1.74$ (i.e., $E = 2.47$ GPa and $\nu = 0.46$). The corresponding global displacement residual $\Theta$ is equal to 0.21 pixel. The values $\mu = 861$ MPa and $\kappa = 1.71$ (i.e., $E = 2.54$ GPa and $\nu = 0.47$) are identified when the element size is equal to 32 pixels. It can be noted that even though the displacement uncertainty is (relatively) high, the identification results are consistent thanks to the large number of measurement points. Furthermore, the two measurement techniques yield equivalent results. The latter are also in good agreement with measured properties in tension (i.e., $E = 2.46 \pm 0.02$ GPa and $\nu = 0.45 \pm 0.03$).

## 4. Towards integrated approaches

The results derived in Section 2.1 allow us to propose an integrated approach to the measurement *and* the identification problems. By choosing *a priori* a basis of functions $\varphi_i(\mathbf{x})$ relevant to an experiment, the identification of the unknown levels $\mathbf{v}_i$ not only provides an information on the displacement features but also yields quantities to be identified from a mechanical point of view. This type of technique was used to decompose the displacement field with linear functions, sines and cosines [59] for 1D signals. The same type of procedure was extended to 2D situations in which a spectral decomposition is assumed [60] or to measure the toughness of brittle materials [61].

In the present case, the previous example is analysed again by using an integrated approach. Since the basis function is richer, the number of ZOIs is reduced (here to one) and its size is larger (in the present case up to $l = 1000$ pixels). The chosen basis consists in rigid body motions (translation and rotation, i.e., three degrees of freedom $v_1$, $v_2$ and $v_3$)

$$\mathbf{v}_t = v_1 + iv_2 \text{ and } \mathbf{v}_r = v_3 iz \tag{19}$$

where a complex notation is used ($v = v_x + iv_y$, $z = x + iy$), and the reference solution given in Section 3 [equations (17) and (18)]

$$\mathbf{v}_b = a_b \left[ \frac{(\kappa-1)\mathbf{z}}{R} + \frac{R+\mathbf{z}}{R+\mathbf{z}} - \frac{R-\mathbf{z}}{R-\overline{\mathbf{z}}} - \kappa \log\left(\frac{R+\mathbf{z}}{R-\mathbf{z}}\right) - \log\left(\frac{R+\overline{\mathbf{z}}}{R-\overline{\mathbf{z}}}\right) \right] \text{ with } a_b = \frac{FR}{4\pi\mu} \tag{20}$$

These four functions allow us to account for image differences [equation (5)]. For each choice of the elastic parameters, the decomposition can be performed, but more importantly, a global quality factor can be obtained, and even a local map of non-resolved differences [equation (1)]. This feature can in turn be utilised to identify the best material constants in the sense of lowest global error $\Phi$.

In the present case, we note that the loading only comes into play through the amplitude of one single field, $\mathbf{v}_b$. Moreover, as dictated by homogeneity, the force (line density) scaled by the shear modulus appears in the amplitude. Thus, the objective function is *quadratic* in $1/\mu$. The shear modulus is simply read from the amplitude, $a_b$, of the field $\mathbf{v}_b$ [equation (20)]. Similarly, the objective function to be minimised is quadratic in $\kappa/\mu$. This feature can be used to compute accurately the optimum choice for $\kappa$ and $\mu$.

The difficulty of this direct approach is the fact that the overall amplitude of the displacement along the compression axis exceeds 2 pixels, and hence the Taylor expansion is inadequate without using a severe filtering and hence reducing the computation accuracy. This fact becomes apparent when looking at the map of residuals, where one observes a significant increase of the error as one moves away from the centre of the disk. The way to circumvent this difficulty is to correct for the deformed image using a first estimate of the displacement, and then detect the unresolved differences. This process can be repeated until convergence at the desired accuracy. In the test example shown in figure 7, the amplitude of the residual displacement field is reduced to a few $10^{-2}$ pixel after two iterations. For such displacement the Taylor expansion approximation becomes suited without having to resort to any filtering.

By focusing on a centred ZOI of size $1000 \times 1000$ pixels (the radius of the cylinder is 794 pixels), the following values are obtained for $\kappa = 1.75$ (i.e., $\nu = 0.46$) and $\mu = 878$ MPa (i.e., $E = 2.56$ GPa). Let us note that the size of the analysed domain has little impact on these estimates. A much smaller zone of $600 \times 600$ pixels (i.e., about one third of the previous area) yields $\nu = 0.44$ and $E = 2.51$ GPa. As could have been anticipated by the proximity of these elastic properties with those estimated using the post-processing procedure presented in the previous section, the displacement field that is estimated (figure 7) here is extremely close to previous determinations (figure 4). The strain field being a derivative was noted to be much more sensitive to noise than the displacement field. By using the integrated approach, the derivation procedure does not generate such a short wavelength noise. Figure 8 shows the corresponding estimate of the normal strain along the horizontal direction. It is interesting to note that the spatial average of the strain is equal to $1.6 \times 10^{-3}$, in excellent agreement with the previous estimate based on the Q1-determination of the displacement field. However, the spatial modulation of the strain is here clearly seen, and not hidden in the noise. This observation provides a very rewarding result for including more knowledge on the mechanics of the problem in the identification procedure.

The quality of the measured displacement field can be observed in the map of gray level residuals (figure 9) defined in equation (3) by $\sqrt{\Phi(\mathbf{v})}$ and applied to the present case. An RMS value of 16 gray levels is obtained to be compared with 18 gray levels for the Q1-procedure. The very low relative value confirms the overall quality of the measured displacement field and the different values of the elastic constants. These results are a direct validation of the Q1-procedure thanks to the integrated approach.

Figure **10** shows the field of global error as a function of the shear modulus and Poisson's ratio. To make this estimate dimensionless, the relative overall quality $(\Phi[\mathbf{v}_b(\mu,\nu)] - \Phi[\mathbf{v}_b(\mu_{opt},\nu_{opt})])/\Phi[\mathbf{v}_b(\mu_{opt},\nu_{opt})]$ is shown, where the subscript *opt* denotes the optimal values. A well-marked minimum can be observed.

## 5. Summary and perspectives

A general formulation of the measurement of displacement fields is introduced. Three different approaches are derived. The first two are based upon correlation algorithms that lead to either locally (constant) or globally continuous displacements (in the sense of a Q1-finite element interpolation). An uncertainty analysis is performed so that the correlation parameters can be chosen in relation with displacement uncertainties. The third one, the integrated approach, consists in exploiting a priori information on the mechanical behavior at the measurement stage. It allows one to measure a displacement field that is mechanically admissible (e.g., solution to an elasticity problem, here particularised to a Brazilian test).

The analysis of diametrical compression (i.e., Brazilian test) of a thin disk is illustrated. This case is considered as particularly difficult since the picture texture is not the best that can be achieved. The general problem of identifying elastic parameters in a heterogeneous test from the knowledge of full-field displacements is formulated in terms of a minimization over a basis of mechanically significant displacement fields. The first two measurement techniques yield similar results in terms of elastic properties with similar correlation parameters. Even though the displacement uncertainties are not identical, the fact that numerous measurement points are available makes the identification quite robust. The integrated approach is also applied to the same experimental test. The values of the elastic parameters obtained through this approach are in good agreement with those obtained by using an *a posteriori* analysis of the same test and a tensile test performed on the same material.

This integrated approach can also be applied to other situations for which closed-form solutions exist (e.g., plates with holes, notched or cracked samples, Flamant problem) by using for instance Kolossov-Muskhelishvili potentials as illustrated herein. The latter potentials may even be exploited for more general situations where closed-form solutions do not exist and yet an admissible displacement field is searched for. For non-linear constitutive laws or heterogeneous solids, a similar integrated approach may be followed by using numerical tools such as finite element modeling rather than elastic potentials.

## Acknowledgements

The authors wish to thank Prof. Jean Lemaitre for useful discussions and Bumedijen Raka for performing the tensile test reported herein. This work was performed within the research network "*GDR 2519 : Mesure de champs et identification en mécanique des solides.*"

# Figure captions

Figure 1: Correlation parameters $l$ and $\delta$ in a reference picture (left) and a picture (12-bit digitisation, $1024 \times 1280$ pixel resolution) in the deformed state (right) of a polycarbonate sample face (see figure 3) for a classical DIC approach. In the Q1-procedure, where overall displacement continuity is satisfied, only one parameter is relevant (i.e., $l = \delta$).

Figure 2: -a-Standard displacement uncertainty as a function of the prescribed displacement for different ZOI sizes when the separation $\delta = l$ (classical DIC algorithm). -b-Displacement uncertainty vs. ZOI size $l$. Equation (12) is depicted by the solid and dashed lines for the two correlation techniques.

Figure 3: -a-Experimental configuration of a Brazilian test. -b-Schematic of the experiment and corresponding reference picture.

Figure 4: Vertical (a) and horizontal (b) displacement fields expressed in pixels (1 pixel is equal to 98 μm) in a Brazilian test measured by using a Q1-procedure with $l = 16$ pixels.

Figure 5: Map of the residuals in gray levels at convergence for the Q1-procedure with $l = 16$ pixels. The dynamic range of the analysed pictures (figure 1) is 4096 gray levels.

Figure 6: Raw estimate of the normal strain along the horizontal direction computed from the previously determined displacement field (figure 4). Note that this field is here dominated by the measurement uncertainty. The underlying significant strain level is shown in figure 8.

Figure 7: Vertical (a) and horizontal (b) displacement fields expressed in pixels (1 pixel is equal to 98 µm) in a Brazilian test measured by using an integrated approach with $l = 1000$ pixels.

Figure 8: Map of the normal strain along the horizontal direction computed from the above solution (integrated approach). This map is to be compared with the raw estimate shown in figure 6.

Figure 9: Map of the residuals in gray levels at convergence for the integrated approach with $l = 1000$ pixels. The dynamic range of the analysed pictures (figure 1) is 4096 gray levels.

Figure 10: Map of global error $(\Phi[\mathbf{v}_b(\mu,\nu)] - \Phi[\mathbf{v}_b(\mu_{opt},\nu_{opt})])/\Phi[\mathbf{v}_b(\mu_{opt},\nu_{opt})]$ for different values of the shear modulus $\mu$ and Poisson's ratio $\nu$ when the integrated approach with $l = 1000$ pixels is used.

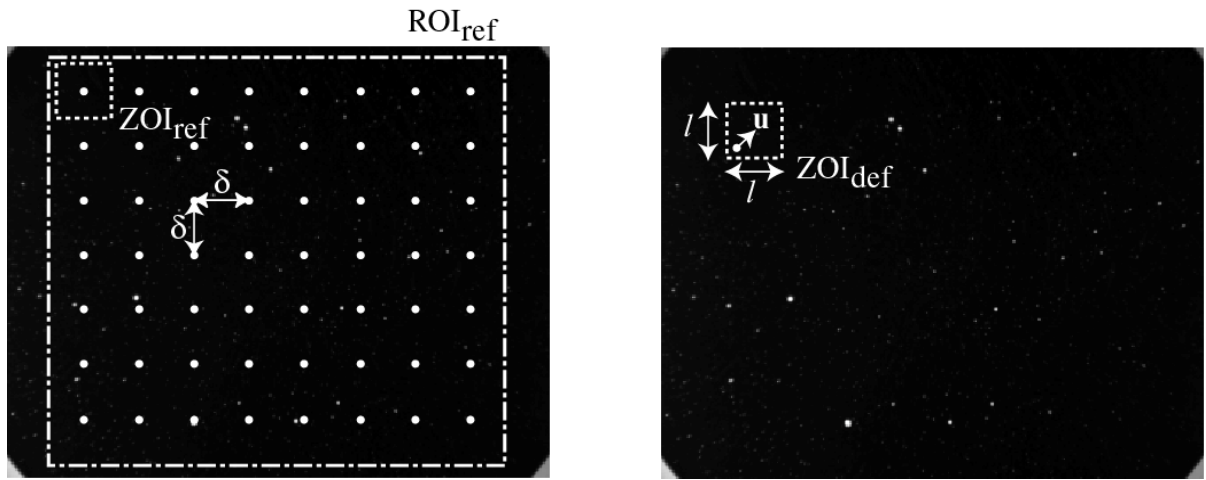

Figure 1: Hild and Roux.

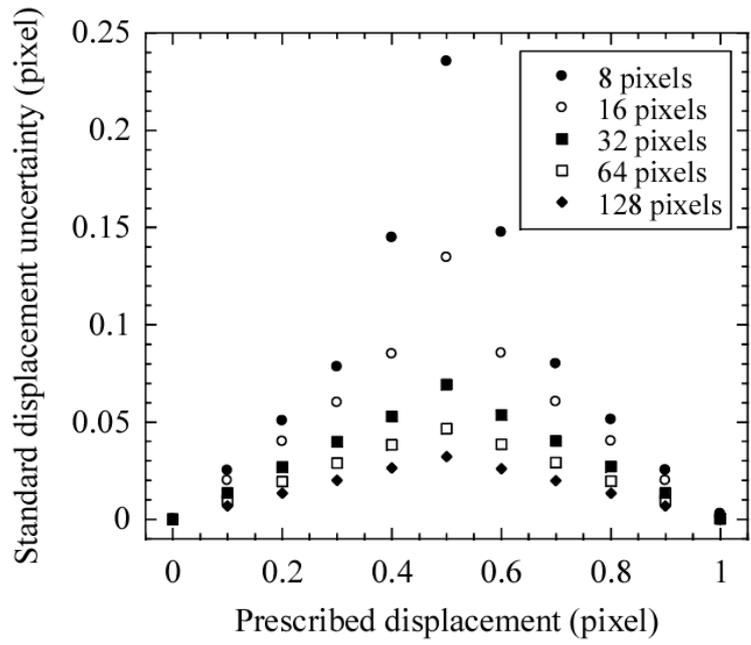

-a-

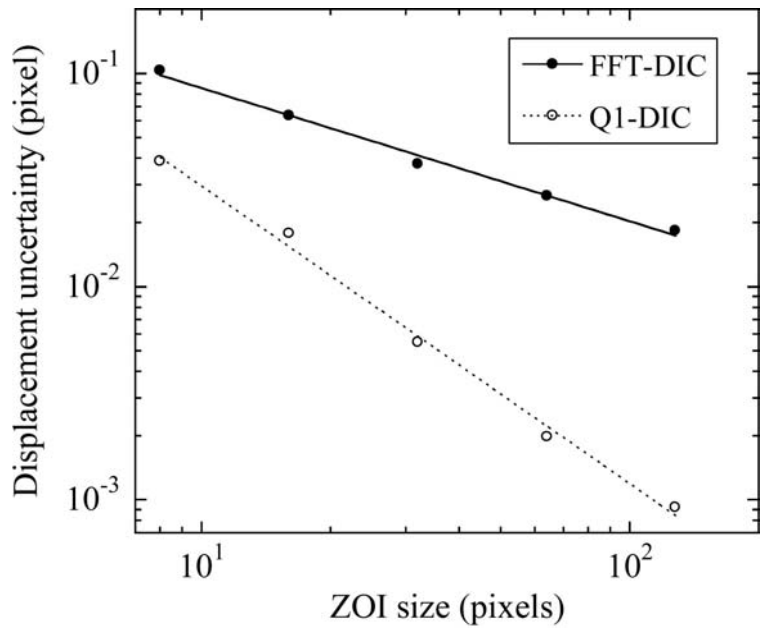

-b-

Figure 2: Hild and Roux.

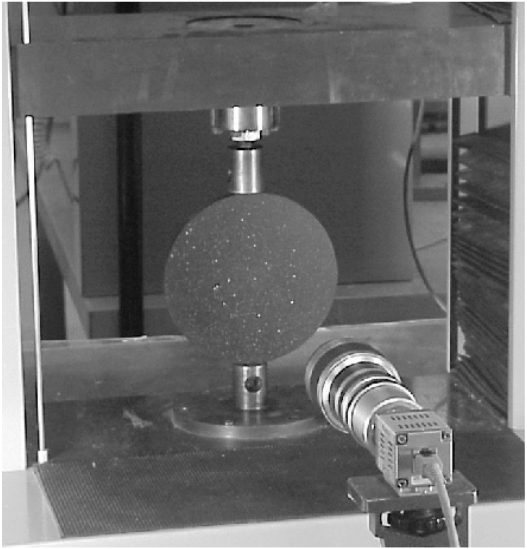    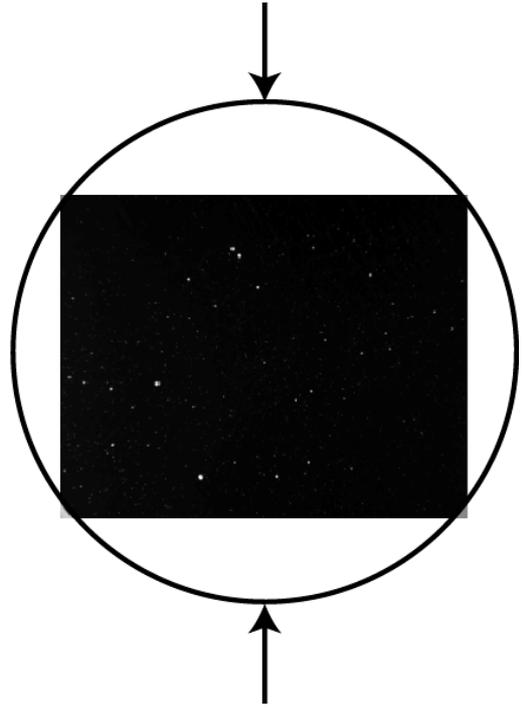

-a-                                        -b-

Figure 3: Hild and Roux.

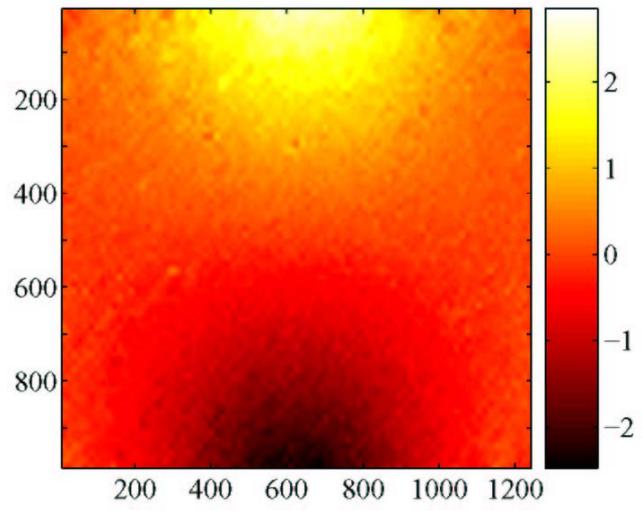

-a-

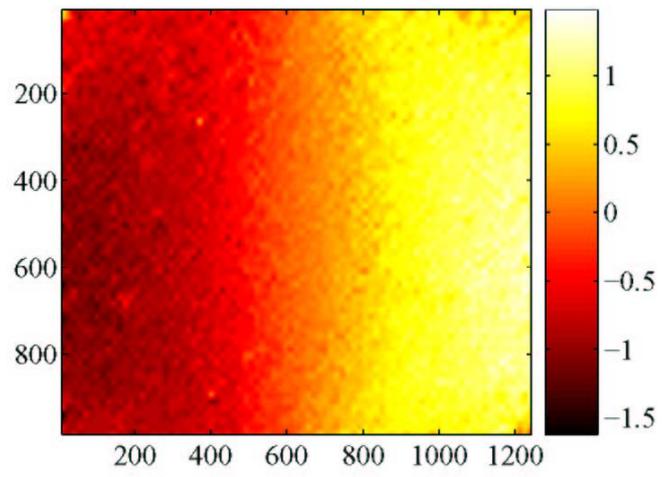

-b-

Figure 4: Hild and Roux.

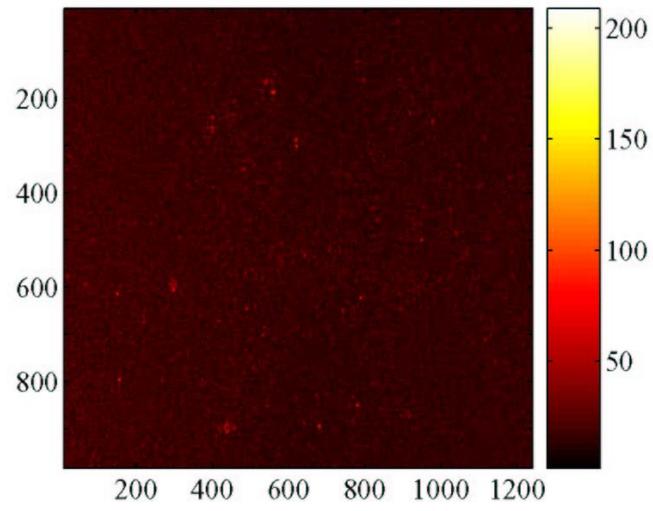

Figure 5: Hild and Roux.

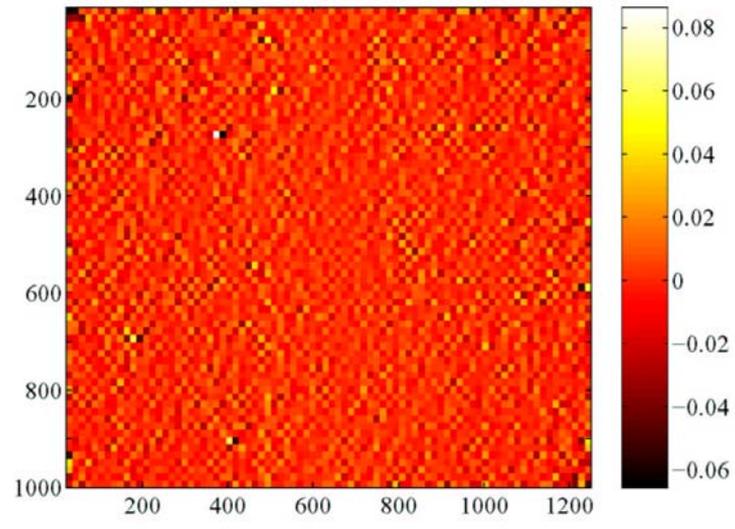

Figure 6: Hild and Roux.

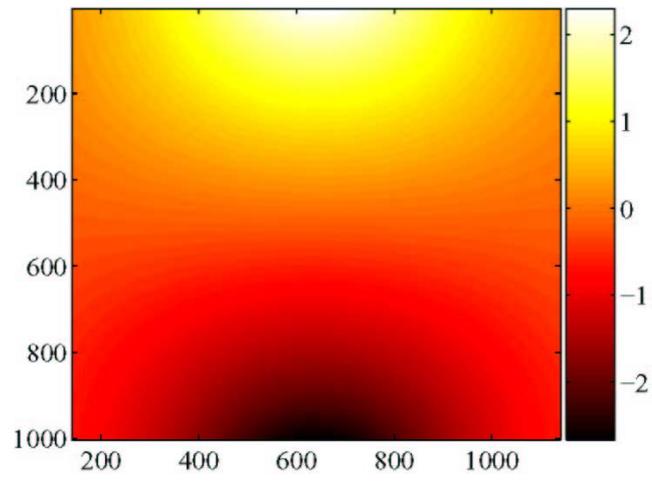

-a-

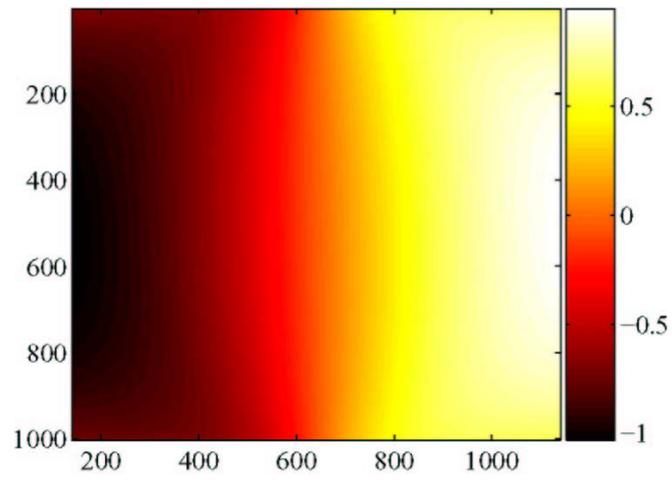

-b-

Figure 7: Hild and Roux.

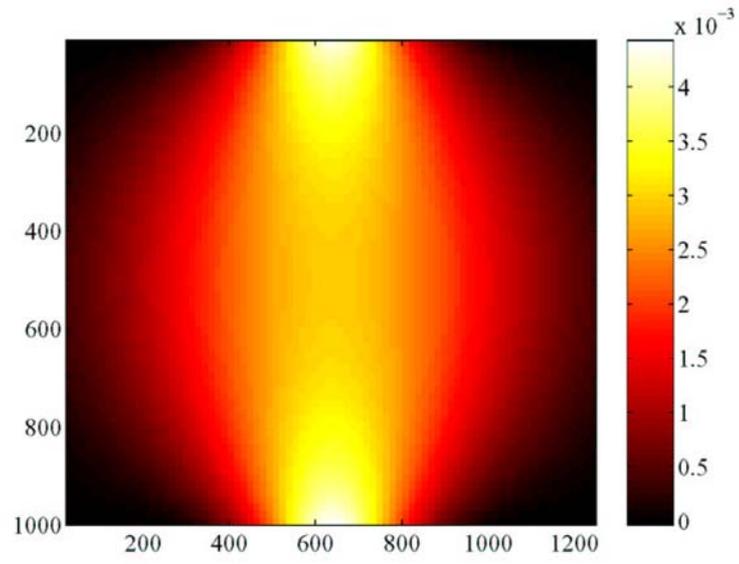

Figure 8: Hild and Roux.

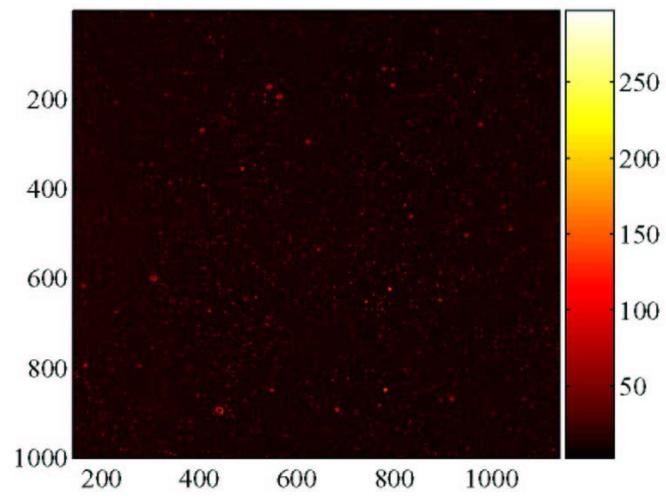

Figure 9: Hild and Roux.

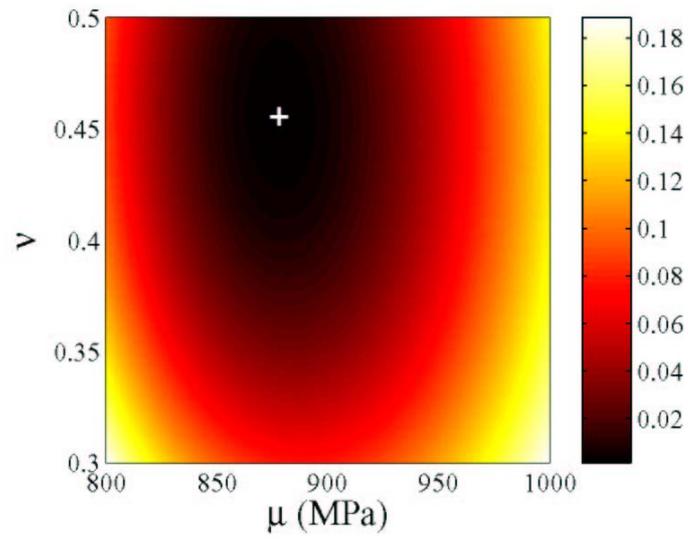

Figure 10: Hild and Roux.